\documentclass[oneside,english]{elsart}
\usepackage{pslatex}
\usepackage[T1]{fontenc}
\usepackage[latin1]{inputenc}
\usepackage{amsmath}
\usepackage{graphicx}
\usepackage[numbers]{natbib}

\makeatletter
\usepackage{natbib}
\usepackage{amssymb}
\date{1st April 2004}

\usepackage{babel}
\makeatother
\begin{document}
\begin{frontmatter}

\title{Spatially Varying Steady State Longitudinal Magnetization in Distant
Dipolar Field-based Sequences}

\author{C. A. Corum}

\address{AHSC, P.O. Box 245067, Tucson, AZ 85724-5067 USA}

\ead{corum@email.arizona.edu}

\ead[url]{http://www.u.arizona.edu/\%7Ecorum}

\author{A. F. Gmitro}

\begin{abstract}
Sequences based on the Distant Dipolar Field (DDF) have shown great
promise for novel spectroscopy and imaging. Unless spatial variation
in the longitudinal magnetization, $M_{z}(s)$, is eliminated by relaxation,
diffusion, or spoiling techniques by the end of a single repetition,
unexpected results can be obtained due to spatial harmonics in the
steady state $M_{z}^{SS}(s)$ profile. This is true even in a homogeneous
single-component sample. We have developed an analytical expression
for the $M_{z}^{SS}(s)$ profile that occurs in DDF sequences when
smearing by diffusion is negligible in the $TR$ period. The expression
has been verified by directly imaging the $M_{z}^{SS}(s)$ profile
after establishing the steady state.
\end{abstract}
\end{frontmatter}

\section{Introduction}

NMR and MRI sequences utilizing the Distant Dipolar Field (DDF) have
the relatively unique property of preparing, utilizing, and leaving
spatially-modulated longitudinal magnetization, $M_{z}(s)$, where
\(\hat{s}\) is in the direction of an applied gradient. In fact this
is fundamental to producing the novel {}``multiple spin echo''\cite{DBD79,BBG90}
or {}``non-linear stimulated echo'' \cite{ASDK97} of the classical
picture and making the {}``intermolecular multiple quantum coherence
(iMQC)'' \cite{QH93} observable in the quantum picture.

Existing analytical signal equations for DDF/iMQC sequences depend
on $M_{z}(s)$ being sinusoidal during the signal build period\cite{alw98b,CG04b}.
Experiments that probe sample structure also require a well-defined
{}``correlation distance'' which is defined as the repetition distance
of $M_{z}(s)$ \cite{RB96,WAM+98,ACM02}. If the repetition time $TR$
of the DDF sequence is such that full relaxation is not allowed to
proceed $TR<5T_{1}$, or diffusion does not average out the modulation,
spatially-modulated longitudinal magnetization will be left at the
end of one iteration of the sequence. The next repetition of the sequence
will begin to establish {}``harmonics'' in what is desired to be
a purely sinusoidal modulation pattern. Eventually a steady state
is established, potentially departing significantly from a pure sinusoid. 

\begin{figure}
\includegraphics[%
  width=5.5in,
  keepaspectratio]{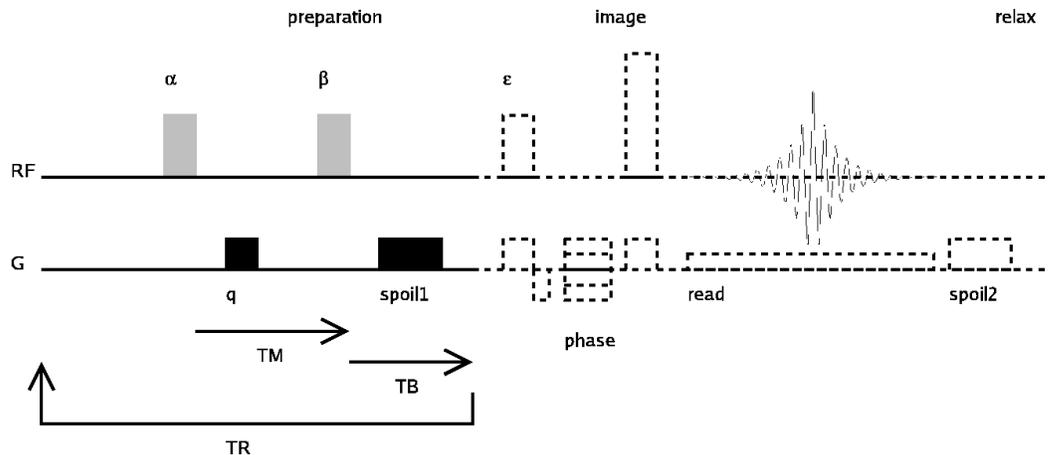}

\caption{\label{fig:Pulse-Sequence}Pulse Sequence. All RF pulses shown as
hard for simplicity are actually Sinc3. $\alpha$ and $\beta$ are
the same phase.}
\end{figure}

\section{Experimental Methods\label{sec:Experimental-Methods}}

In order to study the behavior of the steady state $M_{z}^{SS}(s)$
profile we have implemented a looped DDF preparation subsequence followed
by a standard multiple-phase encode imaging sub-sequence. (Figure
\ref{fig:Pulse-Sequence}.) The $\alpha$ pulse excites the system,
the gradient $G_{q}$ twists the transverse magnetization into a helix.
$\beta$ rotates one component of the helix back into the longitudinal
direction. For simplicity we have omitted the $180^{\circ}$ pulses
used to create a spin echo during TM and/or TB sometimes present in
DDF sequences. Also, we are only interested in $M_{z}(s)$ in this
experiment, not the actual DDF-generated transverse signal. Looping
the {}``preparation'' sub-sequence thus creates the periodic $M_{z}(s)$
profile, spoils remaining transverse magnetization, and establishes
$M_{z}^{SS}(s)$. The $\varepsilon$ pulse converts $M_{z}^{SS}(s)$
into transverse magnetization, allowing it to be imaged via the subsequent
spin echo {}``image'' sub-sequence. $M_{z}^{SS}(s)$ must be re-established
by the {}``preparation'' sub-sequence for each phase encode. After
a suitably long full relaxation delay {}``relax,'' the sequence
is repeated to acquire the next k-space line. This is clearly a slow
acquisition method because many $TR$ periods are required to reach
steady state in the preparation before each k-space line is acquired.
The sequence is intended as a tool to directly image the $M_{z}^{SS}(s)$
profile, verifying the $M_{z}^{SS}(s)$ that would occur in a steady
state DDF sequence, not as a new imaging modality. 

\begin{figure}
\includegraphics[%
  width=5.5in,
  keepaspectratio]{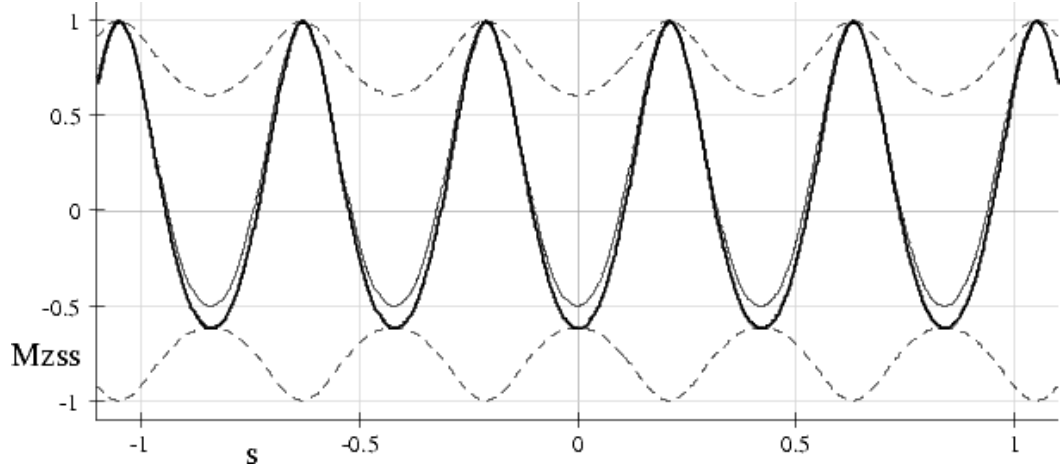}

\caption{\label{fig:Mz_theory_image}Theoretical values of $M_{z}(s)$. $M_{z}^{SS}(s)$is
shown dashed $---$ as an envelope, $M_{z}^{SS,\,\beta}(s)$ is shown
as a heavy line, $M_{z}^{SS,\, TB}(s)$ as a normal line. $\alpha=\beta=90^{\circ},\  TR=2s,\  TM=0ms,\  TB=100ms,\ \  T_{1}=1.4s$}
\end{figure}

\section{Theory}

The effect of the ''preparation'' pulse sequence was first determined
for a single iteration. The progress along the sequence is denoted
by the the superscript.

Starting with fully relaxed equilibrium magnetization before the $\alpha$
pulse:\begin{equation}
M_{z}^{Eq}(s)=M_{0}\label{eq:initial}\end{equation}
after the $\alpha$ pulse, the mix delay $TM$ and the $\beta$ pulse
we have:\begin{align}
M_{z}^{\beta}(s) & =[A^{\beta}cos(q\, s)+B^{\beta}]\, M_{z}^{Eq}+C^{\beta}M_{0}\label{eq:beta}\end{align}
\[
A^{\beta}=-sin(\alpha)\, e^{-\frac{TM}{T_{2}}}sin(\beta)\]
\[
B^{\beta}=cos(\alpha)\, e^{-\frac{TM}{T_{1}}}cos(\beta)\]
\[
C^{\beta}=(1-e^{-\frac{TM}{T_{1}}})\, cos(\beta)\]
The parameter $q=\frac{2\pi}{\lambda}$, where $\lambda$ is the helix
pitch resulting from the applied gradient. Diffusion has been assumed
to be negligible at the scale of $\lambda$. Note that $T_{2}$ is
used in $A$ rather than $T_{2}^{*}$ when $G_{q}$ is larger than
background inhomogeneity and susceptibility gradients.

After the build delay $TB$ we have:

\begin{equation}
M_{z}^{TB}(s)=[A^{TB}cos(q\, s)+B^{TB}]\, M_{z}^{Eq}(s)+C^{TB}M_{0}\label{eq:TE}\end{equation}
\[
A^{TB}=-sin(\alpha)\, e^{-\frac{TM}{T_{2}}}sin(\beta)\, e^{-\frac{TB}{T_{1}}}\]
\[
B^{TB}=cos(\alpha)\, e^{-\frac{TB}{T_{1}}}cos(\beta)\, e^{-\frac{TB}{T_{1}}}\]
\[
C^{TB}=[(1-e^{-\frac{TB}{T_{1}}})\, cos(\beta)-1]\, e^{-\frac{TB}{T_{1}}}+1\]

At the start of the next repetition, after a $TR$ period inclusive
of $TM$ and $TB$ we have\begin{equation}
M_{z}^{TR}(s)=[A^{TR}cos(q\, s)+B^{TR}]M_{z}^{Eq}(s)+C^{TR}\, M_{0}\label{eq:one}\end{equation}
\[
A^{TR}=-sin(\alpha)\, e^{-\frac{TM}{T_{2}}}sin(\beta)\, e^{-\frac{TR-TM}{T_{1}}}\]
\[
B^{TR}=cos(\alpha)\, cos(\beta)\, e^{-\frac{TR}{T_{1}}}\]
\[
C^{TR}=[(1-e^{-\frac{TM}{T_{1}}})\, cos(\beta)-1]\, e^{-\frac{TR-TM}{T_{1}}}+1\]

If we apply the sequence $N$ times and re-arrange the terms we get
the series:\begin{equation}
M_{z}^{NxTR}(s)=M_{0}+M_{0\,}[A^{TR}cos(q\, s)+B^{TR}+C^{TR}-1]\sum\limits _{n=1}^{N}[A^{TR}cos(q\, s)+B^{TR}]^{n-1}\label{eq:sum}\end{equation}
for the starting magnetization state after $N$ repetitions of the
sequence.

Summing an infinite number of terms results in the expression for
the steady state $M_{z}^{SS}(s)$ after a large number of TR periods:\begin{equation}
M_{z}^{SS}(s)=M_{0}-M_{0\,}[\frac{A^{TR}cos(q\, s)+B^{TR}+C^{TR}-1}{A^{TR}cos(q\, s)+B^{TR}-1}]\label{eq:steady_state}\end{equation}

One can then calculate the magnetization state after the $\beta$
pulse in the steady state:\begin{equation}
M_{z}^{SS,\,\beta}(s)=[A^{\beta}cos(q\, s)+B^{\beta}]\, M_{z}^{SS}(s)+C^{\beta}M_{0}\label{eq:ss_beta}\end{equation}

and after $TB$:

\begin{equation}
M_{z}^{SS,\, TB}(s)=[A^{TB}cos(q\, s)+B^{TB}]\, M_{z}^{SS}(s)+C^{TB}M_{0}\label{eq:ss_tb}\end{equation}

We show graphs of Eq. {[}\ref{eq:steady_state}{]}, {[}\ref{eq:ss_beta}{]},
and {[}\ref{eq:ss_tb}{]} in Figure \ref{fig:Mz_theory_image} for
$TR=2s$.

\begin{figure}
\includegraphics[%
  width=5.5in,
  keepaspectratio]{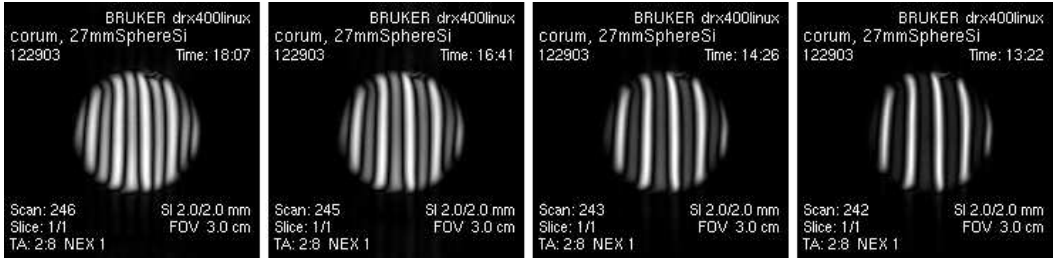}

\caption{\label{fig:Mz_images}$M_{z}^{SS}(s)$ images, $TR=5s,\ 2s,\ 1s,\ 500ms$
from left to right. $TM=TB=~7ms,\  relax=10s$.}
\end{figure}

\section{Results}

\begin{figure}
\includegraphics[%
  width=5.5in,
  keepaspectratio]{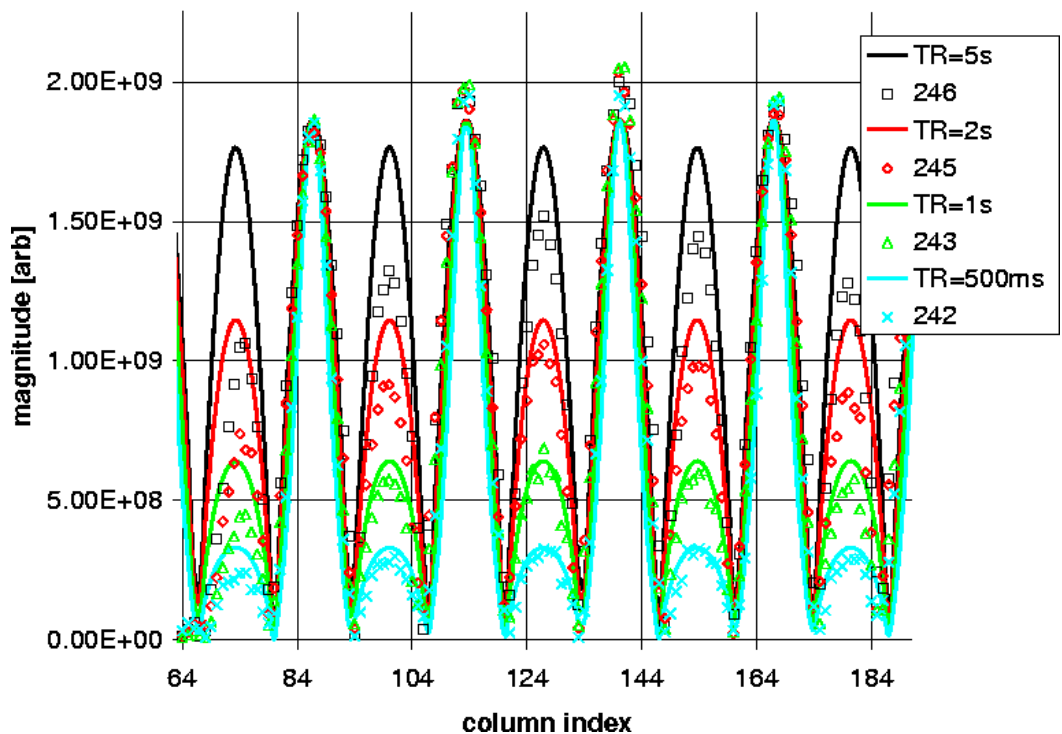}

\caption{\label{fig:Mz_fit}Row 128 data (points) and Fit (lines), $\alpha=\beta=90^{\circ},\  TR=2s,\  TM=TB=7ms,\ \  T_{1}=1.4s\  relax=10s$. }
\end{figure}

We now show in Figure \ref{fig:Mz_images} representative $M_{z}^{SS}(s)$
magnitude images obtained with the sequence described in section \ref{sec:Experimental-Methods}
for four different values of $TR=5s,\ 2s,\ 1s,\ 500ms$. Figure \ref{fig:Mz_fit}
shows several cross sections through row \#128 of Figure \ref{fig:Mz_images}.
The object is an 18mm glass sphere filled with silicone oil. Data
points are superimposed with the corresponding magnitude of the theoretical
curve. The $T_{1}$ of the silicone oil (at 400MHz) was measured by
spectroscopic inversion recovery to be 1.4s. A Bruker DRX400 Micro
2.5 system was used with a custom 27mm diameter 31P/1H birdcage coil.
10 $TR$ periods were used to establish steady state. A 10s {}``relax''
delay was used between phase encodes to establish full relaxation.
$G_{q}$ was 3ms and 2.5mT/mm, with $G_{spoil1}$ of 5ms and 100mT/mm.
No attempt was made to account for $B_{1}$ inhomogeneity. A single
scaling parameter was used for all theoretical curves. We achieved
good agreement with the theoretical predictions. In the sequence as
used $TM=TB=~7ms$. A variety of other $G_{q}$ directions and strengths
show similar agreement with theory. Better agreement in the fit between
experiment and theory can be obtained with $\alpha=\beta=75^{\circ}$than
with the nominal $90^{\circ}$. A $B_{1}$ map needs to be determined
to see if this corresponds more closely to the actual experimental
conditions.

\section{Conclusions}

The expressions developed and verified above should be useful to those
wishing to understand or utilize harmonics in the $M_{z}^{SS}(s)$
profile in DDF based sequences in the situation where the diffusion
distance during $TR$ compared with $\lambda$ in negligible. This
is especially true for those carrying out structural measurements
which depend on a well defined correlation distance. The theory should
also hold for spatially varying magnetization density $M_{0}=M_{0}(\vec{r})$,
and longitudinal relaxation $T_{1}=T_{1}(\vec{r})$.

\section{Acknowledgements}

This work and preparation leading to it was carried out under the
support of the Flinn Foundation, a State of Arizona Prop. 301 Imaging
Fellowship, and NIH 5R24CA083148-05.

\bibliographystyle{elsart-num}
\addcontentsline{toc}{section}{\refname}\bibliography{mz}

\end{document}